# Are temperature reconstructions regionally biased?

O. Bothe[1]

**Abstract**

Are temperature reconstructions possibly biased due to regionally differing density of utilized proxy-networks? This question is assessed utilizing a simple process-based forward model of tree growth in the virtual reality of two simulations of the climate of the last millennium with different amplitude of solar forcing variations. The pseudo-tree ring series cluster in high latitudes of the northern hemisphere and east Asia. Only weak biases are found for the full network. However, for a strong solar forcing amplitude the high latitudes indicate a warmer first half of the last millennium while mid-latitudes and Asia were slightly colder than the extratropical hemispheric average. Reconstruction skill is weak or non-existent for two simple reconstruction schemes, and comparison of virtual reality target and reconstructions reveals strong deficiencies. The temporal resolution of the proxies has an influence on the reconstruction task and results are sensitive to the construction of the proxy-network. Existing regional temperature biases can be attenuated or accentuated by the skill of the reconstruction approach.

---
1
CliSAP Universität Hamburg and Max-Planck-Institute for Meteorology
Email: oliver.bothe [ at ] zmaw.de

## 1. Introduction

Paleo proxy records give evidence of a climatic distinct period in medieval times. Most reconstructions and paleoclimate simulations indicate warmer conditions in the northern hemisphere extratropics for this period with temperatures reaching levels of mid or even end 20$^{th}$ century warmth (see e.g. Diaz et al., 2011, Graham et al., 2011, Jungclaus et al., 2010, Feulner, 2011). However, the timing of this warm period or climatic anomaly is uncertain. Reconstructions of forcing agents (e.g. Crowley, 2000, Schmidt et al., 2011 and references therein) hint to the possibility that a cluster of volcanic eruptions culminating in the 1258 eruption of unknown location in concert with the onset of the so called Wolf Minimum in solar activity produced a radiative disturbance large enough to interrupt or even terminate the medieval warm and climatic anomalous period. The forcing reconstructions further indicate a solar minimum (Oorth minimum) in the early 11$^{th}$ century with preceding high but unexceptional solar activity and some persistent but as well not exceptionally intense volcanism at the change of the millennia. Reconstructions of CO2 content partially agree with these forcing assumptions (Jungclaus et al., 2010, Frank et al., 2010).

Temperature reconstructions display frequently (Ljungqvist, 2010; Esper et al, 2002; Christiansen and Ljungqvist, 2012) a peak warmth in the period 950 to 1050 at levels of northern hemisphere temperature not experienced again until the 20$^{th}$ century and only a small response to the reduced solar activity in the Oorth Minimum. On the other hand, simulations of the climate of the last Millennium generally display maximum warmth in the 12$^{th}$ century (e.g. Jansen et al., 2007, Ammann et al., 2007; Jungclaus et al., 2010) during the reconstructed medieval solar maximum subsequent to the Oorth minimum. Thus, the simulated medieval warming is generally in agreement with the evolution and the amplitude of the applied forcing agents (solar, volcanic, land use change, e.g., Schmidt et al., 2011; Crowley, 2000, Jansen et al., 2007), but not necessarily with the temperature reconstructions.

As some proxy networks consider a higher number of high latitude proxies, it may be asked whether this leads to a bias in the temperature estimates. Such biases could be due to reductions in summer insolation due to changes in orbital parameters for high latitudes as found in simulation studies (Fischer and Jungclaus, 2010; Servonnat et al., 2010). This in turn leads to changes in the seasonal cycle and to variations in the seasonal contrast between high and mid latitudes. This orbital effect could have resulted in warmer high latitude annual and winter temperature anomalies at the start of the last millennium compared to today. Jones et al. (2003) comment on the importance of changes in the annual cycle for the biogeochemical proxy-properties.

For this study, it is also of interest that one recent reconstruction (Christiansen and Ljungqvist 2012, CL12) includes a high percentage of east Asian proxies. Contrasting to the possible orbital effects in high latitudes, there is no clear indication for a biasing effect of east Asian proxies. However, in interpreting east Asian climate proxies some peculiarities have to be considered as for example the importance of the Tibetan Plateau as a source of elevated atmospheric heating and the relation of the (east) Asian summer monsoon to Pacific decadal variability (e.g. Chang 2000) and tropical Pacific SST-variability (e.g. Wang 2000).

For the sake of argument of this note I assume, that science gives two estimates of the climate in the early second millennium of the common era: (i) models indicate an northern hemisphere extratropical climate evolution with a warm period in medieval times peaking in the $12^{th}$ century; (ii) reconstructions hint to a millennium with maximum warmth at its beginning and its end. Based on the VSLite model for artifical tree rings (Tolwinski-Ward, 2011) and the community simulations of the last millennium (Jungclaus et al., 2010), I shortly discuss whether the choice of location may drive the reconstructed temperature series towards an over-estimation of the temperature at the beginning of the millennium. This is not meant to criticize the work or the choices made in the reconstructions, but rather to identify possible pitfalls. Note that, for example, CL12 clearly show that the regional selection of proxies has only a weak influence on the reconstruction.

## 2. Data and methods

The VSLite algorithm of Tolwinski-Ward et al. (2011) is utilized to compute pseudo-tree growth series for data of two of the community simulations of the climate of the last millennium performed at the Max Planck Institute for Meteorology (Jungclaus et al., 2010), one with weak and one with stronger amplitude of solar forcing variations. Tree growth was simulated for 108 permutations of four growth parameters (maximum and minimum thresholds of growth for moisture and temperature) over all grid points (land and ocean) between 30N and 75N. For the present study, an ensemble average over these 108 sets is used.

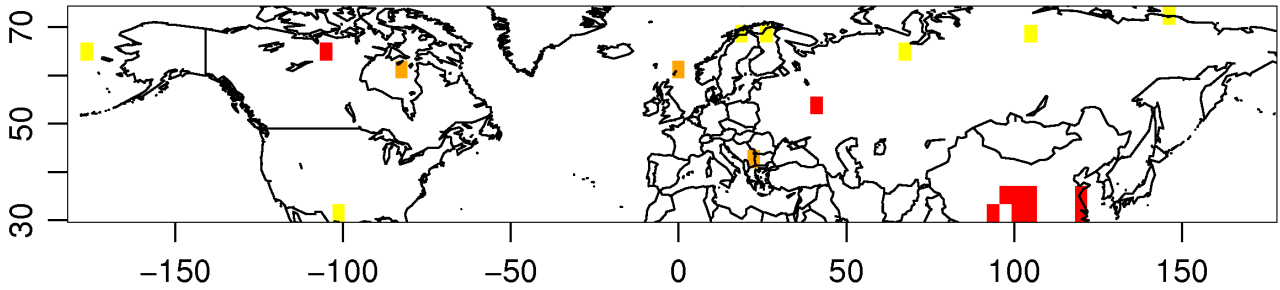

*Figure 1: Pseudo tree ring locations used in the optimized set. Yellow locations are annually resolved, orange ones 5 year running means and red ones 11 year running means.*

I use two sets of 16 proxy locations (Figure 1). Each set has nine series in high latitudes north of about 60 degrees and of the remaining seven four represent the eastern half of China from the Tibetan Plateau to the Yangtze river delta. This is similar to the network of CL12, and the temporal resolutions of the pseudo proxies are adjusted as well. The first set (set 1) of pseudo proxies is optimized to include only locations which correlate with local temperatures of both simulations and with the hemispheric extratropical target of both simulations (average 30N to 90N). I only present results for this network. For another set (set 2), the nine high latitude records are located north of 65N with used grid points approximating the locations of CL12. However, I replace Dye-3, GISP2 and Reland (compare CL12) as no pseudo-treerings are computed over the Greenland icecap or long periods are represented by no growth (not shown). The "optimization" of set 1 implies that temperature and pseudo growth-series may be uncorrelated for set 2. Results are generally worse for set 2. Figure 1 displays the grid boxes selected for set 1.

Two simple reconstruction approaches are applied, a composite plus scaling version and a local approach following Christiansen (2010) and Christiansen and Ljungqvist (2011, 2012). Temperature and proxies are prepared by running means as highlighted in Figure 1. Asian proxies are decadally smooth, as are data for a location in Russia and one high latitude proxy. Five year smooth data is used in two higher latitude locations and a south-eastern European proxy. Otherwise the temporal resolution is annual. Thus high latitude proxies include more noise than the Asian ones. Series are centred and calibrated over the period 1881 to 1960. The reconstructions are validated over the period 1961 to 1990. For both approaches the proxies are standardised by their standard deviation over the full period and multiplied by their correlation to the calibration data (either local or hemispheric). Averaging provides a composite series for the composite approach. A classical calibration is performed which is equivalent to an inverse regression. The resultant is scaled to match the variability of the target. For the local approach, a local inverse regression is performed, the resultant matched to the calibration period temperature variability, and then the local resultants are simply averaged. A cosine weight only slightly alters the reconstruction results. If not stated otherwise, all considered data are annual mean series.

## 3. Results

*3.1. Biased temperatures?*

Figure 2 displays the mean differences between individual grid points and northern hemispheric extratropical mean anomalies (with respect to 1881 to 1960) for the period 1100 to 1200 together with their zonal means for both simulations. A positive deviation is found in the high latitudes reaching up to 0.7K in the strong solar forcing simulation, while anomalies are slightly negative for northern monsoons. The weak solar forcing simulation displays a further small positive difference in mid-latitudes. Thus, the strong solar forcing simulation basically displays an Arctic amplification pattern for this period.

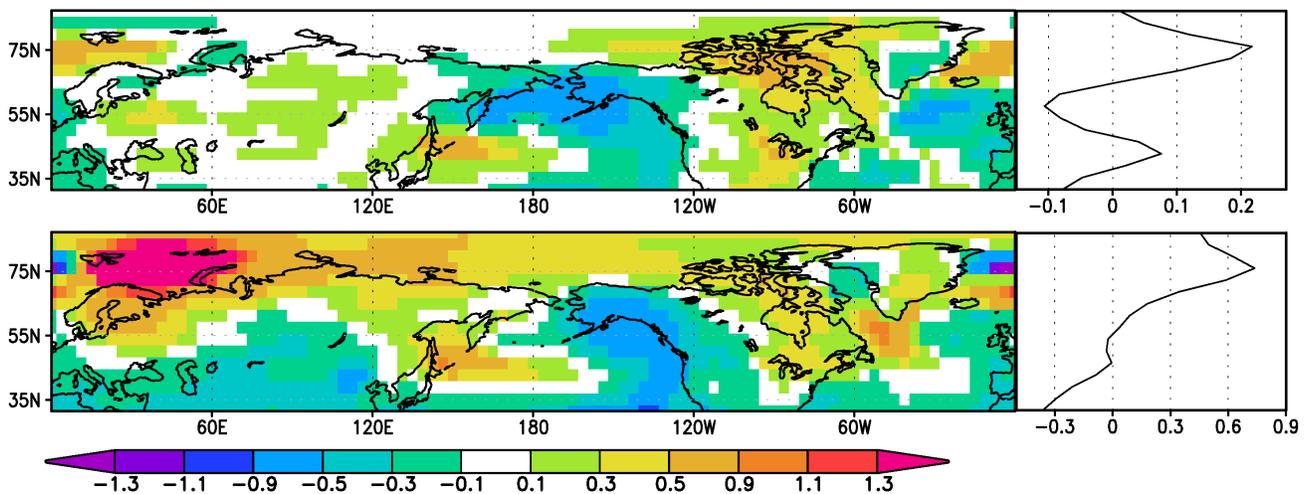

*Figure 2: Mean over the period 1100 to 1200 of differences between grid point annual temperature anomalies and hemispheric extratropical anomaly with respect to the period 1881 to 1960, upper row for weak solar forcing simulation, lower row for strong solar forcing simulation. Right panels are zonal means. Note different scale of right panels.*

For the strong forcing simulation, this results in a slight warm offset of the 51 year moving mean of network average local temperatures compared to the target prior to 1400 AD, which peaks at values larger 0.1K in the 12$^{th}$ and 13$^{th}$ centuries and reaches similar values for the late 19$^{th}$ century (compare Figure 3a). However, for the weak forcing simulation the network average local temperature basically fits the target mean and the moving mean offset varies between about minus 0.09 and plus 0.07K throughout the 1200 year period (compare Figure 3b). Leave one out ensembles generally do not differ much from the full local estimate. Under-estimation of temperature anomalies is notable for the weak forcing simulation.

The warm offset becomes more pronounced for both simulations, if Asian locations are excluded (orange curve in Figure 3a,b) or only high latitude locations (red curve in Figure 3a,b) are considered. For the weak forcing simulation, this is mostly an expression of an enhanced variability with prominent cold excursions, which is particularly pronounced for the high-latitude only curve. On the other hand exclusion of high-latitude locations (blue curve in Figure 3a,b) or considering only Asian locations (magenta curve in Figure 3a,b) underestimates lower frequent variability for the weak forcing simulations and mostly misses the warm excursions of hemispheric temperature for both simulations. This is obvious for the Asia only series, which also is colder than other estimates over much of the simulation period for the strong forcing simulation. Figure 2 already indicates the possibility of such a cold bias for Asia in this simulation. If the network is not correlation optimized, the positive deviations increase and a warm offset is particularly strong for the weak forcing simulation in the 10$^{th}$ and leading into the 11$^{th}$ century. While the results are inconclusive, they emphasize the possible pitfall of regional biases due to regional over- and under-representation.

*3.2. Biased reconstructions?*

*a) Reconstruction merit*

How do simple inverse regression based reconstructions succeed under such circumstances (for the "optimized" set see Figures 3c-f,4)? Considering the composite plus scale approach, calibration period reduction of error coefficients (RE, Cook et al., 1994) are negative for both simulations and all 21 reconstructions, while validation period REs are mostly positive though smaller 0.5 for the weak forcing simulation and mostly scattering between zero and 0.1 for the strong forcing simulation (Figure 4a). Coefficients of efficiency are generally smaller -1 for the strong forcing simulation and scatter about 0.3 for the weak simulation (Figure 4d). Thus, reconstruction skill is small for the weak solar forcing simulation environment and non-existent for the strong solar forcing environment. Correlations are always positive, but Pearsons $R^2$ (Figure 4b) are smaller 0.2 (0.1) for the strong forcing simulation in the validation (calibration) period and mostly between 0.3 and 0.45 for the weak forcing simulation and the validation period (calibration period $R^2$ smaller 0.15). Mean offsets from the target are negative and larger an absolute value of 0.2 K for the strong forcing simulation and mostly positive and scattering about 0.1 for the weak forcing simulation (Figure 4c).

Without "optimizing" the proxy locations, the most prominent differences for the composite approach are weakened (slightly increased) correlations and REs for the strong (weak forcing simulation) in the validation period, as well as more pronounced mean offsets and worse CEs for the strong forcing simulation (not shown). Measures of merit are worse for the local approach, but merit remains generally better for the weak forcing simulation than for the strong simulation.

*b) Discussion of reconstruction series*

Although calibration period measures (e.g. Res in Figure 4a) are unskillful for both simulations the reconstructions indicate some skill over the validation period (Figure 4). It appears, that reconstructions are better for the weak forcing simulation and best for the composite plus scaling approach. Figure 3c-f plot the related reconstructed time series and the target time series. Obviously both approaches have problems with the task.

The composite approach gives very good agreement with the target for the full set of proxies in the strong forcing reality. Leave one out and high latitude reconstructions underestimate the low frequency variability. Most notably, reconstructed warm periods are less warm than in the simulated reality. Asia and mid-latitude sub-sets overestimate low frequency variability, which leads to pronounced cold but also some overly warm excursions.

Underestimation of low frequent variability is the most common feature of the local approach reconstructions and influences the reconstruction of simulated warm episodes over the last millennium in the strong forcing simulation. Most sub-sets scatter around the full reconstruction. The Asia only sub-set prominently overestimates cold excursions since the 13[th] century.

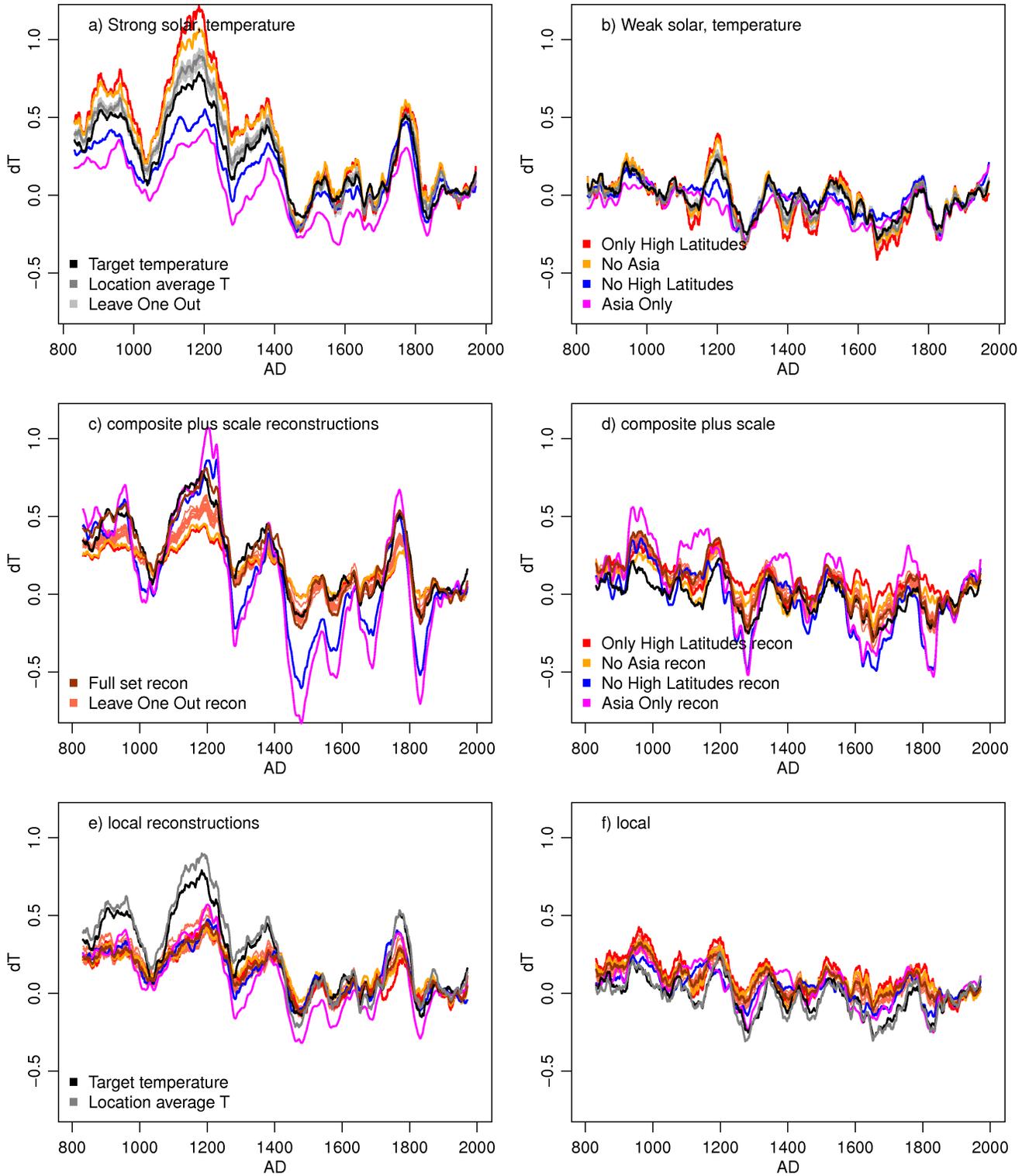

*Figure 3: Northern hemispheric mean temperatures (a,b) and reconstructions (c-f) based on different location sets (see legends) and simulations (left, strong solar full forcing, right, weak solar full forcing.*

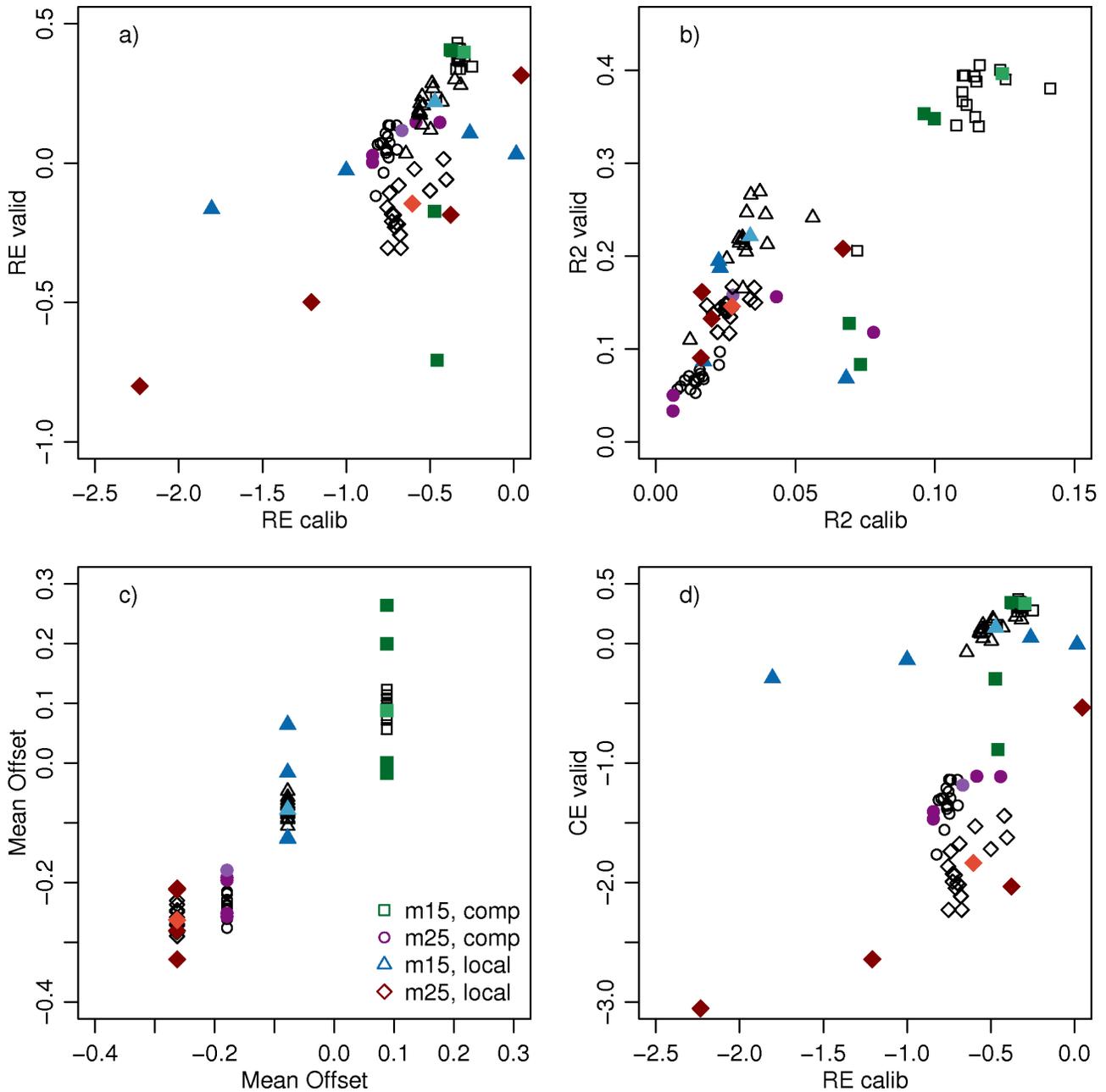

*Figure 4: Measures of merit for the different reconstructions (see legend for symbols and colors). Filled symbols for sub-sets of locations (light colors, full set) and open symbols for leave one out sets of all different reconstructions.*

For the weak solar forcing simulation, the composite approach and the full network result in a reconstruction with a pronounced warm bias over much of the last millennium. Leave one out ensembles scatter around the full reconstruction. Subsets emphasizing the high latitudes are also generally warmer than the target, display less low frequent variability and underestimate the cold 17[th] century conditions. Sub-sets excluding the high latitudes overestimate low frequent variability, the amplitude of warm and cold excursions and the duration of warm episodes.

The full local reconstruction is as well warm biased compared to the target in the weak forcing environment. Most sub-sets agree in this behaviour. The Asia only subset is slightly closer to the cold episodes of the target, but also notably overestimates the duration of warm episodes.

Without "optimization", the reconstructions generally become worse (not shown). Most prominent differences are a stronger underestimation of low frequent variability for high latitude and leave one out sets for the strong forcing simulation and a stronger warm bias for the weak forcing simulation in the composite approach. The local approach gives in parts better results for this network in the strong forcing simulation in recent centuries. Sub-sets excluding high-latitude locations do not capture low frequent variations and warm temperatures for both simulations. In the weak forcing simulation, most sub-sets again present a warm bias, which is enhanced compared to the optimized set.

Comparing local temperature sub-sets and all the reconstructions indicates the following general problems. The composite approach mainly suffers from resolving too few variability, although the full reconstruction follows the target quite closely. The local approach, on the other hand, appears to be dominated by the Asian low frequent proxies. Reconstructions agree largely with the Asian local sub-set temperature.

## 4. Discussion and conclusions

The present assessment tries to highlight possible pitfalls in reconstructing the climate of the last 1000 years from limited sets of proxies. The emphasis is on networks which are dominated by high latitude proxies. The assessment is certainly influenced by the characteristics of the COSMOS-Mill simulator, the applied forcing series, the statistical characteristics of the computed pseudo tree growth series and the processes considered in computing them with the VSLite algorithm. I willingly use only simple regression based reconstruction approaches and for the main part chose locations where tree growth correlates to both simulation targets and local temperature. In addition, the pseudo growth is very likely warm season biased and the warm season may display the largest high latitude biases over a thousand year period (Briffa and Osborne 2002; Jones et al., 2003). The noisy inter-annual characteristics of the high latitude data reduces their importance in the presented reconstructions, while the decadal resolution of the Asian data possibly overemphasizes the latter locations.

Thus, the network average temperature is indeed in some periods warm biased compared to the extratropical hemispheric target. This is most pronounced in the strong solar forcing simulation from the 9$^{th}$ to the 14$^{th}$ century. Indeed, the high latitude locations are responsible for this bias, whereas lower latitude locations are cooler than the extratropical hemispheric average in the strong solar forcing simulation. In the weak solar forcing simulation, the low latitude location average is less variable on low frequencies thus under-representing warm and cold excursions. However, the evaluation, however, does not show a general warm bias of reconstructions based on high latitudinal climate.

Furthermore, already the calibration period merit is low. Subsequent reconstructions suffer to some extent from a loss of variance due to averaging. Further shortcomings are an overestimation of low frequent variability for non-high latitude sub-sets resulting in overestimating the intensity of warm and cold episodes in the composite approach. Both utilized methods give a pronounced warm bias in the weak forcing simulation. Interestingly, rather good and rather uninformative reconstructed temperature series arise when varying the reconstruction method, the utilized simulation forcing or the pseudo networks (keeping the same proportion of high-latitude to non-high latitude locations and of east Asian to non-east-Asian locations).

Even though the calibration merit is low, which would have possibly prohibited a real world reconstruction, I conclude from the presented results the following: Millennial temperature reconstructions may well be regionally biased. Warm biases may indeed occur over a thousand year period for chronologies which rely heavily on high latitudes. Additionally the well known problem of variance loss due to averaging becomes obvious once more, while on the other hand an exaggerated variance also arises as a possibility. That is, besides of regional biases, the temporal resolution of the data has an impact on the resultant reconstructed series, and the mixing of temporal resolutions has to be rigorously assessed. Finally, regional biases may be prominent but the lack of power of the reconstruction methods may either accentuate or attenuate their influence.

*Acknowledgements*
OB is supported through the Cluster of Excellence 'CliSAP' (EXC177), University of Hamburg, funded through the German Science Foundation (DFG).